\documentclass[a4paper, 12pt, openbib ]{article}
\usepackage{times}
\usepackage{pifont}
\usepackage{amsmath}
\usepackage{amssymb}
\usepackage{amsfonts}
\usepackage{graphicx}
\usepackage{floatflt}
\usepackage{geometry}
\usepackage{layout}
\usepackage{enumerate}
\usepackage{bm}
\usepackage{tikz}
\usetikzlibrary{patterns}
\def\ner{\boldsymbol}
\usepackage{amsthm}
\usepackage{thmtools}

\def\tfract#1/#2{{\textstyle{\raise0.8pt\hbox{$\scriptstyle#1$}\over%
\hbox{\lower0.8pt\hbox{$\scriptstyle#2$}}}}}
\def\mezzo{\tfract 1/2 }

\def\quarto{\tfract 1/4 }

\def\radi2k{\tfract 1/{\sqrt {2k}} }
\def\der{\partial }
\def\cvd{\vbox{\hrule \hbox to 9 pt {\vrule height 9 pt \hfil \vrule} \hrule}}
\def\downnormalfill{$\,\,\vrule depth4pt width0.4pt
\leaders\vrule depth 0pt height0.4pt\hfill\vrule depth4pt width0.4pt\,\,$}
\def\WT#1{\mathop{\vbox{\ialign{##\crcr\noalign{\kern3pt}
      \downnormalfill\crcr\noalign{\kern0.8pt\nointerlineskip}
      $\hfil\displaystyle{#1}\hfil$\crcr}}}\limits}
\def\be{\begin{equation}}
\def\ee{\end{equation}}
\def\bes{\begin{equation*}}
\def\ees{\end{equation*}}
\def\bea{\begin{eqnarray}}
\def\eea{\end{eqnarray}}
\def\beas{\begin{eqnarray*}}
\def\eeas{\end{eqnarray*}}
\def\ba{\begin{array}{rcl}}
\def\ea{\end{array}}
\def\der{\partial}
\numberwithin{equation}{section}
\def\go{\leavevmode \raise.3ex\hbox{$\scriptscriptstyle \langle\!\langle\!  $}%
~\ignorespaces}
\def\gf{\relax \ifhmode \unskip~\else \leavevmode \fi \raise.3ex\hbox{$\! \scriptscriptstyle\rangle\!\rangle\, $}}
\title{
{\Large  Quasi-particles, thermodynamic  consistency \\ and the gap equation}
{\vskip 0.6 truecm}} 

\author{{\large  Enore~Guadagnini} \\  {\normalsize {~}} \\  {\normalsize  Dipartimento di Fisica {\it E. Fermi} dell'Universit\`a di Pisa,} \\  {\normalsize and INFN Sezione di Pisa,} \\ {\normalsize   Largo B. Pontecorvo  2, 56127 Pisa, Italy.} 
}
\date{}

\begin{document}

\maketitle 

\vskip 0.7 truecm 

\begin{abstract}

The thermodynamic properties of  superconducting electrons are usually studied by means of the quasi-particles distribution; but in this approach, the ground state energy and the dependence of the chemical potential on the electron density cannot be determined. In order to solve these problems,  the thermodynamic potentials are derived by means of the Bogoliubov-Valatin formalism. The thermodynamic potentials  can be   obtained by computing the free energy of a  gas of  quasi-particles, whose energy spectrum is conditional on  the gap function. However, the nontrivial dependence of the gap   on the temperature jeopardises  the validity of the standard  thermodynamic relations.  In this article it is shown how the thermodynamic consistency ({\it i.e.} the validity of the Maxwell  relations) is recovered, and  the correction terms to the quasi-particles potentials  are computed.   
It is shown that the Bogoliubov-Valatin transformation avoids the problem of the thermodynamic consistency of the quasi-particle approach; in facts, the correct identification of the variables, which are associated with the quasi-particles, leads to a precise calculation  of the quasi-particles vacuum energy  and  of the dependence of the chemical potential on the electron density. 
The stationarity condition for the grand potential coincides with the gap equation, which   guarantees  the  thermodynamic consistency.  The expressions of various   thermodynamic potentials, as functions of the  $(T,V,N)$ variables,  are produced in the low temperature limit; as a final check, a rederivation of the condensation energy is presented.  
 \end{abstract}

\vskip 1.2 truecm

\section{Introduction}
\label{intro}

The Bogoliubov transformation is one of the main discoveries in mathematical physics. In this article it is shown how to compute the thermodynamic potentials ---functions of the $ (T,V,N)$ variables--- in superconductivity by means of the Bogoliubov transformation. 
It is demonstrated that the Bogoliubov transformation avoids the problem of the thermodynamic consistency of the quasi-particle approach.  The quasi-particles vacuum energy and the dependence of the chemical potential on the electron density are derived. 

In the low temperature limit, the thermodynamic behaviour of any macroscopic physical system  is determined by the structure of the energy levels of the system  which belong to a neighbourhood of the ground state energy.  The experimental data show that, usually,   these energy levels are organized  in such a way that one can give a quite good description of their thermodynamic properties 
by means of a gas of quasi-particles \cite{L1,L2}. The energy spectrum of a single quasi-particle is specified by its dispersion relation, which  is determined by the nature of the interactions among the ``atoms'' or the ``elementary constituents'' of the system.   

Let us consider the framework in which the gas of quasi-particles describes the thermal fluctuations of the system  above its ground state. In this formalism, the number of quasi-particles is not conserved ---in facts it vanishes at zero temperature--- and therefore the quasi-particles chemical potential is zero. 
 Even if the system is composed of strongly interacting atoms ---like in a quantum liquid--- the quasi-particles are weakly interacting, and then one can determine, with a good approximation,  the thermodynamic   potentials of the system. Indeed, in the low temperature limit the density of quasi-particles becomes very low, so that their mutual interactions produce minor effects. Thus a crucial part of the  information on the thermal properties of the whole system is mainly contained in the form of the dispersion relation of a single quasi-particle. Let $\varepsilon (\ner p)$ be the energy of a single quasi-particle in the state labelled by the value $\ner p$ of the momentum.  The function $\varepsilon (\ner p)$ can be a rather complicated function of the momentum, which can be determined by the experimental data, or it can be deduced ---in some approximated form--- from the hamiltonian of the system. In superconductors, for instance, the dispersion relation for a single   quasi-particle \cite{BCS,L3} assumes the form 
\be
\varepsilon (\ner p) \simeq \left [ \left (  \ner p^2 / 2m -\mu \right)^2 + \Delta_{\ner p}^2   \right ]^{1/2} \; , 
\label{1.1}
\ee
where $\mu$ denotes the chemical potential for the electrons. The gap function $\Delta_{\ner p}$  can be different from zero in a neighbourhood of the surface of the Fermi sphere and satisfies 
\be
\Delta_{\ner {p}} = \left  \{ \begin{array}  {l@{ ~ } l}   \Delta (T) & {\; \; } \hbox{ when } \left | \, {  \ner {p}^2 / 2 m } - \mu \right | \leq  \delta E  \simeq \hbar \omega_D\, ;   \\  ~ & ~ \\  
~~~~~0 &  {\; \; } \hbox{ otherwise; }  
\end{array} \right. 
\label{1.2}
\ee
in which $\omega_D$ indicates the Debye frequency of the ions lattice. The gap $\Delta (T)  $  is not vanishing when the temperature $T$ takes values below the critical value $T_c$, and $\Delta (T) =0 $ for $T \ge T_c$.    As a result, the energy $\varepsilon $ of a single quasi-particle nontrivially depends on the temperature, $\varepsilon = \varepsilon ( \ner p , T) $. But when  the dispersion relation  $\varepsilon ( \ner p , T) $ nontrivially depends on the temperature, the standard expressions of the  thermodynamic  potentials ---derived in statistical mechanics--- for a gas of weakly interacting  constituents do not satisfy the thermodynamic relations (or Maxwell relations). 

More precisely, if $A$ represents a standard thermodynamic potential which is computed  in statistical mechanics when the atomic  energy $\varepsilon_0 (\ner p) $ does not depend on the temperature, let us denote by $A_q$ the corresponding potential for the gas of quasi-particles  which  is obtained from $A$ by means of the substitution $\varepsilon_0 (\ner p) \rightarrow \varepsilon (\ner p , T)$.  For instance, the free energy $F_q$ and the internal energy $U_q$ of a gas of noninteracting quasi-particles  with energy $\varepsilon (\ner p , T)$, satisfying the Fermi-Dirac statistics, are given by  
\be
F_q(T,V) = - {kT V g\over h^3} \int d^3p  \, \ln \left ( 1 + e^{- \varepsilon ( \ner p , T) / kT} \right ) \; ,  
\label{1.3}
\ee
and 
{\vskip - 0.8 cm}
\be
U_q (T,V) =  {V g \over h^3 } \int d^3p \, {\varepsilon ( \ner p , T)  \over e^{ \varepsilon ( \ner p , T) / kT} + 1}      \; \, , 
\label{1.4}
\ee
where $g$ denotes the spin degeneration factor. When the thermodynamic relations are satisfied, the entropy $S$ is given by $S = - \der F(T,V)/ \der T$; and since $F = U - TS$, one has $U = F - T (\der F / \der T)$. However, if $[\der \varepsilon (\ner p , T) / \der T ] \not= 0 $, 
one finds  
\be
U_q \not=  F_q - T \left ( {\der F_q \over \der T} \right )_V \; ,  
\label{1.5}
\ee
and therefore expressions (\ref{1.3}) and (\ref{1.4}) do not satisfy the thermodynamic relations. 

 In addition to their agreement with  experiments,  the thermodynamic relations codify to some extent the laws of thermodynamics and are  necessary for the sake of logic.  The requirement of validity of the thermodynamic relations will be called the  thermodynamic consistency. The thermodynamic consistency implies that, in general,  the thermodynamic potentials  of the macroscopic systems cannot be exactly equal to the potentials $A_q$ of the quasi-particles.  The presence of adjusting terms $\delta A_q$ is needed in order to rectify the $A_q$ expressions and make  the final combinations  $A_q + \delta A_q$  correct.  As far as the  computation of $\delta A_q$ is concerned, the knowledge of the quasi-particles distribution is of no help, because $\delta A_q$ is essentially determined by the quasi-particles vacuum energy.  
 
 One of the  purposes of the present article  is to  produce the correction terms to the quasi-particles potentials in the case of superconductivity by means of the Bololiubov-Valatin formalism. It is shown that the accurate  determination of the quasi-particles vacuum energy completes the construction of the thermodynamic potentials  satisfying the Maxwell relations. It is  verified that, in agreement with the Landau principle,  the value of the gap corresponds to a stationary point of the grand potential $\Omega (T,V,\mu )$. Then the expressions of the thermodynamically consistent potentials as  functions of the $(T,V,N)$ variables are obtained, and how the chemical potential is related with the electron density is  determined. As a final check, a derivation  of the superconducting condensation energy from $\Omega (T,V,\mu )$ is presented. 
 
The thermodynamic consistency in the presence of nontrivial medium-dependent dispersion relation has been considered for instance by Shanenko,  Yujalova and Yukalov \cite{SYY,SYYZ} in the case of clustering matter,  and by Gorenstein and Yang \cite{GY} in the case of a gluon plasma. This issue has been elaborated also in \cite{BST,YS,BA,GF1,GF2}. The Landau principle, which can be  interpreted as an equilibrium condition for ordinary thermal states, implies the thermodynamic consistency. It as been argued \cite{GF2} that, conversely,  the validity of the thermodynamic consistency does not necessarily imply  the Landau principle. The results for the superconducting electrons are in  agreement with the analysis of Ref.\cite{GY}. 

In superconductivity, the dependence of the thermodynamic potentials on the statistical mean value $N$ of the number of electrons is a rather nontrivial issue  because, when $T < T_c$, the $U(1)$ symmetry which is related to the number  of  electrons in superconductors  is broken \cite{BCS,L3}.  The  solution to this problem which  is presented in the following chapters and the outcomes of the computed thermodynamic potential $\Omega (T,V,\mu )$ 
are in  agreement with the known results \cite{BCS,L3,TK,REW} on superconductivity. 
In the present work it is shown that  the  identification of the variables which are associated with the quasi-particles leads to a precise determination of the quasi-particles vacuum energy.  The method, which is presented here to solve this question,  can find applications  also in the study of  topological superconductivity and in the new developments on the topological states of matter \cite{TAE,MAE,KAN,QZ,DRMO,NDC,LC,KB,NBK,A}.   

The basic concepts  which are connected with the use of the  quasi-particles  distribution are briefly recalled in Section~{\ref{sec:2}}.  Starting from the BCS hamiltonian, in Section~{\ref{sec:3}} the Bogoliubov-Valatin formalism \cite{NNB,VA} is used to determine the grand potential $\Omega $,   according to the procedure envisaged for instance by Rickayzen  \cite{RIC}. By means of a Bogoliubov transformation  acting on  the creation and annihilation operators of the  electron   fluctuations around the Fermi sphere,   the full expression of $\Omega (T, V, \mu ) $ satisfying the Maxwell relations is  obtained.    The thermodynamic consistency of the result  is discussed in Section~{\ref{sec:4}}, where a new derivation of  the gap equation is described.   In Section~{\ref{sec:5}}  the relation connecting the chemical potential with the electron density is derived,  and  the consistent expressions of various thermodynamic potentials ---functions of the $(T,V,N)$ variables---  of the superconduting electrons  are produced in the low temperature limit.  

\section{Quasi-particles distribution}
\label{sec:2}

 Let us briefly recall the basic notions  which are related to the use of  the quasi-particles formalism.  The  description of the low temperature physics by means of quasi-particles represents a phenomenological approach in which the dispersion relation of a single quasi-particle nontrivially depends on the macroscopic variables of the material $\varepsilon (\ner p) = \varepsilon (\ner p , T, \mu , ... ) $.  Thus, in addition to the temperature,  the failure of the thermodynamic consistency for the quasi-particle potentials $A_q$ really concerns several  state variables.  In certain circumstances the dependence of $\varepsilon $ on these variables is  weak enough so that, in a limited range of variability, one can assume \cite{PN} that $\varepsilon$ only depends on the momentum.    But in general ---and in particular when a phase transition occurs--- one cannot neglect the dependence of $\varepsilon $ on the macroscopic  variables; this dependence will be denoted  by $\varepsilon = \varepsilon (\ner p , T)$. 
  
As it has been suggested by Landau \cite{L2},  a possible way  to connect the thermodynamic potentials of the system with the quasi-particles functions makes use of the quasi-particles distribution.   The only thermodynamic potential whose absolute value is fixed ---and cannot be modified by any additive constant--- is the entropy $S$,  
\be
S = k \, \ln \bigl \{  \hbox{number of microstates with fixed macro conditions}  \bigr \} \; . 
\label{2.1}
\ee
Since the quantum states of the system are described (in the low temperature  limit) precisely  by the gas of quasi-particles,   the entropy of the system is equal to the entropy of the quasi-particles gas. For noninteracting quasi-particles, the number of microstates in a given macrostate can be  determined by means of the distribution $n_{\ner p}$  of the quasi-particles. Indeed, in the quantum case of Fermi statistics, the  analogue of the Boltzmann $H$-functional is given  \cite{L2} by
\be
S  = - {Vgk\over h^3} \int d^3p \, \left [ n_{\ner p} \ln n_{\ner p} + (1-n_{\ner p}) \ln (1-n_{\ner p})\right ] \; . 
\label{2.2}
\ee
Let us recall that, in the case of particles with vanishing chemical potential, the entropy represents the thermodynamic potential for the variables $(U, V )$. This means that, with fixed $(U, V )$, the stable thermal state of the system corresponds to a maximum of $S$. Therefore the distribution $n_{\ner p}$ can be determined by the requirement that, for fixed $(U, V )$,   the variation of $S$ with respect to a generic fluctuation $\delta n_{\ner p }$ must vanish. The associated (conditioned) variational principle   takes the form 
\be
{\delta S \over \delta n_{\ner p} } - \lambda  
{\delta U \over \delta n_{\ner p} } = 0 \; , 
\label{2.3}
 \ee
where the value of the Lagrange multiplier $\lambda $ is given by $\lambda = 1/ T$ as a consequence of the Maxwell relation $TdS -dU -P dV =0$. Because of the explicit dependence (\ref{2.2}) of $S$ on $V$, there is no need to introduce a Lagrange multiplier for the volume. Since a modification of $n_{\ner p}$ causes the following  change in the energy 
\be
\delta U = {Vg\over h^3} \int d^3p \, \varepsilon (\ner p , T)\, \delta n_{\ner p}  \; ,  
\label{2.4}
\ee
the solution of equation (\ref{2.3}) is given precisely  by the Fermi-Dirac distribution 
\be 
 n_{\ner p} = {1 \over e^{\varepsilon (\ner p , T) / kT} + 1 } \; . 
\label{2.5}
\ee
By using equation (\ref{2.5}) several variables can be computed.  Unfortunately, the knowledge of the distribution (\ref{2.5}) ---or of $S(T,V)$---  alone is not enough to determine all the potentials; for instance, the complete expressions of the free energy $F$ and of the internal energy $U$ cannot be obtained  from equation (\ref{2.5}). 

\section{Computation of the grand potential}
\label{sec:3}

The low energy BCS hamiltonian \cite{BCS} for conducting electrons in superconductor metals can be written as   
\be
H_{BCS} = \sum_{\ner {p} , s } {\ner p^2 \over 2 m} \,  b^\dagger_{\ner {p} , s }  b_{\ner {p} , s } + \sum_{\ner {p} , \ner {q} } U_{\ner {p} , \ner {q} } \, b^\dagger_{\ner {p} , + }  b^\dagger_{- \ner {p} , - }  b_{- \ner {q} , - }  b_{\ner {q} , + } \; , 
\label{3.1}
\ee
where  $b^\dagger_{\ner p , s}$ and $b_{\ner p , s}$ denote the creation and annihilation operators for one electron in the state $| \ner p , s \rangle $, where $s= \pm$ refers to the value of one component of the spin,   
\bea
\{ b_{\ner {p} , s} \, , b^\dagger_{\ner {q} , r } \} &=& \delta_{sr} {h^3\over V} \delta (\ner {p} - \ner {q})\quad , \nonumber \\  
\{ b_{\ner {p} , s} \, , b_{\ner {q}, r } \} &=& 0 = \{ b^\dagger_{\ner {p} , s} \, , b^\dagger_{\ner {q} , r } \} \; . 
\label{3.2}
\eea
The interaction kernel $U_{\ner {p} , \ner {q}}$   is related to the amplitude of the  electron-electron  scattering. $U_{\ner {p} , \ner {q}}$  does not depend on the values of the electron thermodynamic variables, as the temperature $T$, the volume $V$ and the chemical potential $\mu $. The main effects of the  interactions between  electrons in the superconducting materials  ---which are relevant for the superconducting phase transition---   are found when $ \left |{  \ner {p}^2 / 2 m } - \mu \right | \leq   \hbar \omega_D $ and $ \left |{  \ner {q}^2 / 2 m } - \mu \right | \leq   \hbar \omega_D$, and in this region $U_{\ner {p} , \ner {q}}$   can be approximated by $- U_0/ V$, where $U_0$ is a positive constant.  The grand partition function $\cal Q$, which   is defined by 
\be
{\cal Q}(T,V,\mu ) =  {\rm Tr} \left [ e^{- H_{BCS} / kT  } e^{N\mu / kT } \, \right ] \equiv  {\rm Tr} \left [ e^{- H  / kT  } \, \right ] \; , 
\label{3.3}
\ee
 can be interpreted as the partition function of a system with total hamiltonian 
\be
H =  \sum_{\ner {p} , s } \eta ( p)  \,  b^\dagger_{\ner {p} , s }  b_{\ner {p} , s } + \sum_{\ner {p} , \ner {q} } U_{\ner {p} , \ner {q} } \, b^\dagger_{\ner {p} , + }  b^\dagger_{- \ner {p} , - }  b_{- \ner {q} , - }  b_{\ner {q} , + } \; , 
\label{3.4}
\ee
in which the ``effective'' kinetic energy $\eta(p )$ of one electron (where $p = | \ner p |$) is given by 
\be
\eta ( p) =  {  \ner {p}^2 \over 2 m } - \mu \; .   
 \label{3.5}
\ee
 Since the trace (\ref{3.3}) cannot be evaluated  exactly, one can introduce a self-consistent  approximation (some kind of a mean field approximation)  which permits to proceed with the computation. In the grand canonical ensemble, let us introduce the statistical mean values 
\be
X_{\ner {p}} = \langle b_{- \ner {p} , - }  b_{\ner {p} , + } \rangle \quad , \quad 
X^*_{\ner {p}} = \langle b^\dagger_{ \ner {p} , + }  b^\dagger_{- \ner {p} , - } \rangle \; .  
\label{3.6}
\ee
In order to simplify the exposition, let us assume that $X^*_{\ner {p}} = X_{\ner {p}} $ (in facts, it turns out \cite{BCS,L3,TK} that one can always choose the phases of the creation and annihilation operators in such a way that the mean values (\ref{3.6}) are real). The couple of operators $b^\dagger_{\ner {p} , + }  b^\dagger_{- \ner {p} , - }$ and $b_{- \ner {q}, - }  b_{ \ner {q} , + }$ can be written as the sum of their means values plus a fluctuation term,  
\bea
b^\dagger_{\ner {p} , + }  b^\dagger_{- \ner {p} , - } &=& X_{\ner {p}} + ( b^\dagger_{\ner {p} , + }  b^\dagger_{- \ner {p} , - } - X_{\ner {p}} ) \; , \nonumber \\
b_{- \ner {q}, - }  b_{ \ner {q} , + } &=& X_{\ner {q}} + ( b_{- \ner {q}, - }  b_{ \ner {q} , + } - X_{\ner {q}} ) \; .   
\label{3.7}
\eea
Now one can assume \cite{TK,RIC} that the effects of the fluctuations are sufficiently small so that,  inserting the identities (\ref{3.7}) into the expression  (\ref{3.4}), the resulting term which is quadratic in the fluctuations can be neglected.  One then finds 
\be
H  \simeq  \sum_{\ner {p} , s } \eta ( p)  \,  b^\dagger_{\ner {p} , s }  b_{\ner {p} , s }  -   \sum_{\ner {p} } \Delta_{\ner {p}  }  X_{\ner {p}} 
  + \sum_{\ner {p}} \left [ 
\Delta_{\ner p } \, b^\dagger_{\ner {p}, + }  b^\dagger_{- \ner {p}, - } 
+ \Delta_{\ner p} \, b_{- \ner {p} , - }  b_{ \ner {p}, + } \right ]    \; ,  
\label{3.8}
\ee
where 
\be
\Delta_{\ner {p}} = \sum_{\ner {q} } U_{\ner {p} , \ner {q} } X_{\ner {q}} \; .  
\label{3.9}
\ee
In order to recover the quasi-particles description of the low temperature behaviour of the system, it is convenient to introduce  new creation and annihilation operators. The corresponding   procedure is composed of two steps: 

\begin{enumerate}

\item introduction of the operators  $d^\dagger_{\ner p , s}$ and $d_{\ner p , s}$ which create and annihilate quasi-particles in the case of free electrons; 

\item introduction of the operators $a^\dagger_{\ner p , s}$ and $a_{\ner p , s}$   ---which are obtained from $d^\dagger_{\ner p , s}$ and $d_{\ner p , s}$ by means of a Bogoliubov transformation--- that diagonalize  the hamiltonian (\ref{3.8}).  

\end{enumerate}
 
\noindent {STEP 1.} The operators $d^\dagger_{\ner p , s}$ and $d_{\ner p , s}$ should describe the one-particle fluctuations with respect to the configuration of the Fermi sphere. Then $d^\dagger_{\ner p , s}$ and $d_{\ner p , s}$  are chosen in such a way that their vacuum state coincides with the free electrons ground state corresponding to the Fermi sphere.  In the $T \rightarrow 0$ limit, the electron states with $p = | \ner p | < p_0 $ are occupied whereas the states with $p = | \ner p | >  p_0 $ are empty, where  
\be
p_0 = \sqrt {2 m \mu \, } \; . 
\label{3.10}
\ee
Therefore $d^\dagger_{\ner p , s}$ and $d_{\ner p , s}$ are defined \cite{NNB} by 
\bea
d^\dagger_{{\ner p} , s} &=&  \left  \{ \begin{array}  {l@{ ~ } l}  
 b_{- {\ner p}, - s }  & \quad \hbox{ for } p < p_0  \; ;    \\   
 %~ & ~ \\  
b^\dagger_{{\ner p}, s }   &  \quad \hbox{ for } p> p_0 \; ;  
\end{array} \right. \quad , \nonumber \\ 
d_{{\ner p} , s} &=&  \left  \{ \begin{array}  {l@{ ~ } l}  
 b^\dagger_{- {\ner p}, - s }  & \quad \hbox{ for } p < p_0  \; ;    \\   
 %~ & ~ \\  
b_{{\ner p}, s }   &  \quad \hbox{ for } p> p_0 \; ,  
\end{array} \right.
\label{3.11}
\eea
The creation operator  $d^\dagger_{{\ner p} , s}$ of a quasi-particle corresponds to the  creation of a one-electron hole in the Fermi sphere if $| \ner p | < p_0$, and  the creation of a one-electron occupied state if $| \ner p | > p_0$.  As a result of the creation of a hole in the Fermi sphere, which corresponds to an electron state with momentum $- \ner p$ and spin component $- s$,  the total momentum of the system increases by $\ner p$ and the total spin of the system increases by $s$. For this reason, when $ p < p_0 $ the quantum numbers of the operators $d^\dagger_{\ner p , s}$ and $d_{\ner p , s}$ are opposite with respect to the quantum numbers of $b^\dagger_{\ner p , s}$ and $b_{\ner p , s}$. The operators $d^\dagger_{\ner p , s}$ and $d_{\ner p , s}$ satisfy the canonical anticommutation relations 
\bea
\{ d_{\ner {p} , s} \, , d^\dagger_{\ner {q} , r } \} &=& \delta_{sr} {h^3\over V} \delta (\ner {p} - \ner {q})\quad , \nonumber \\  
\{ d_{\ner {p} , s} \, , d_{\ner {q}, r } \} &=& 0 = \{ d^\dagger_{\ner {p} , s} \, , d^\dagger_{\ner {q} , r } \} \; . 
\label{3.12}
\eea
The operator  number of quasi-particles $\widehat N_{qp}$ is given by $\widehat N_{qp}= \sum_{\ner p , s}d^\dagger_{\ner {p} , s } d_{\ner {p} , s } $; $\widehat N_{qp}$ counts the  number of holes inside the Fermi sphere plus the number of occupied states outside the Fermi sphere. The number of quasi-particles is not conserved, thus quasi-particles have vanishing chemical potential. In the $T \rightarrow 0$ limit, the number of quasi-particles vanishes.  By using the relation $d_{\ner {p} , s}  d^\dagger_{\ner {p} , s } = 1 - d^\dagger_{\ner {p} , s}  d_{\ner {p} , s }$  and the definition (\ref{3.11}),  one finds  
\be 
   \sum_{\ner {p} , s } \eta ( p)  \,  b^\dagger_{\ner {p} , s }  b_{\ner {p} , s } 
 = \sum_{\ner {p},  s } w( p)  \,  d^\dagger_{\ner {p} , s } d_{\ner {p} , s }    + E_F(V, \mu ) \; , 
 \label{3.13}
 \ee
 in which 
 \be
w( p) = | \eta ( p) | = \left |  {  \ner {p}^2 \over 2 m } - \mu     \right | \; , 
\label{3.14}
\ee
and
\bea
E_F(V , \mu ) &=& - \sum_{\ner {p} , s }^{p < p_0}   w( p) = - 2 \sum_{\ner {p}  }^{p < p_0}   w( p)   \nonumber \\ 
&=& - {2 V  (2m)^{3/2}\over 15 \pi^2 \hbar^3} \mu^{5/2}    \; .
\label{3.15}
\eea
The gas of quasi-particles, which are  associated with the operators (\ref{3.11}), can also be used to compute the thermodynamic variables for a degenerate Fermi gas  (as an alternative to  the Sommerfeld expansion). Note that, differently from the effective kinetic energy $\eta (p)$ of one electron, the kinetic energy $w(p)$ of one quasi-particle satisfies $w(p) \geq 0$. 

\medskip

\noindent {STEP 2.} By using the operators $d^\dagger_{\ner p , s}$ and $d_{\ner p , s}$, the  hamiltonian (\ref{3.8}) reads 
\bea
H  &\simeq&  \sum_{\ner {p} , s } w ( p)  \,  d^\dagger_{\ner {p} , s }  d_{\ner {p} , s } + \sum_{\ner {p}} \left [ 
\Delta_{\ner p } \, d^\dagger_{\ner {p}, + }  d^\dagger_{- \ner {p}, - } 
+ \Delta_{\ner p} \, d_{- \ner {p} , - }  d_{ \ner {p}, + } \right ] \nonumber \\  && \quad + E_F(V, \mu )  -   \sum_{\ner {p} } \Delta_{\ner {p}  }  X_{\ner {p}}    \; .  
\label{3.16}
\eea
Since $d^\dagger_{\ner {p} , -}  d_{\ner {p} , - } = 1 - d_{\ner {p} , -}  d^\dagger_{\ner {p} , - } \, $, one can write 
\be
  \sum_{\ner {p} , s } w ( p)  \,  d^\dagger_{\ner {p} , s }  d_{\ner {p} , s } = 
   \sum_{\ner {p}  } w ( p)  \left [  d^\dagger_{\ner {p} , + }  d_{\ner {p} , + } -  
 d_{ - \ner {p} , - }  d^\dagger_{ - \ner {p} , - }    \right ] 
+ \sum_{\ner {p} } w(p) \; .   
 \label{3.17}
 \ee
Thus the operator (\ref{3.16}) can be written in the form  
\be
H  \simeq \sum_{\ner {p}  } D^\dagger_{\ner {p}}  \begin{pmatrix}  w (p) & \Delta_{\ner {p}} \\ \Delta_{\ner {p}} & - w (p) \end{pmatrix} \! D_{\ner {p}}   + E_F(V , \mu )
 -   \sum_{\ner {p} } \Delta_{\ner {p}  }  X_{\ner {p}}  + \sum_{\ner p} w(p) \; , 
 \label{3.18}  
\ee
where 
 \be
D_{\ner {p}} = \begin{pmatrix} d_{\ner {p} , + }\\ d^\dagger_{-\ner {p} , - } \end{pmatrix} \quad , \quad 
D^\dagger_{\ner {p}} = \left ( d^\dagger_{\ner {p} , + } \, , d_{-\ner {p} , - } \right )\; .    
\label{3.19} 
\ee
By means of the diagonalising matrix  $M_{\ner p} \in SU(2)$  that satisfies the relation 
\be
M^\dagger_{\ner {p}} \begin{pmatrix} w (p) & \Delta_{\ner {p}} \\ \Delta_{\ner {p}} &  - w (p)  \end{pmatrix} M_{\ner {p}} = \begin{pmatrix}  \varepsilon ({\ner {p}}) & 0 \\ 0 &  - \varepsilon ({\ner {p}}) \end{pmatrix} \; , 
\label{3.20}
\ee
in which  
\be
\varepsilon ({\ner {p}} ) = \sqrt {w^2(p) +  \Delta^2_{\ner {p}}} \; ,  
\label{3.21}
\ee
one can introduce  new creation and annihilation operators $a^\dagger_{\ner p , s} $ and $a_{\ner p , s}$ according to 
\be
\begin{pmatrix} a_{\ner {p} , + } \\ a^\dagger_{-\ner {p} , - } \end{pmatrix} = M^\dagger_{\ner p} \cdot D_{\ner p}  \quad , \quad 
 \left ( a^\dagger_{\ner {p} , + } \, , a_{-\ner {p} , - } \right ) = D^\dagger_{\ner p} \cdot M_{\ner p} \; .  
\label{3.22}
\ee
One then obtains 
\bea
\sum_{\ner {p}  } D^\dagger_{\ner {p}}  \begin{pmatrix}  w (p) & \Delta_{\ner {p}} \\ \Delta_{\ner {p}} & - w (p) \end{pmatrix} \! D_{\ner {p}} &=&  \sum_{\ner {p}  } \varepsilon(\ner p) \left [  a^\dagger_{\ner {p} , + }  a_{\ner {p} , + } -    a_{ - \ner {p} , - }  a^\dagger_{ - \ner {p} , - }    \right ] \nonumber \\ 
&=&  \sum_{\ner {p}  , s }  \varepsilon ({\ner {p}})   \,  a^\dagger_{\ner {p} , s }  a_{\ner {p} s } - \sum_{\ner {p}  } \varepsilon ({\ner {p}}) \; .   
\label{3.23}
\eea
By using the operators $a^\dagger_{\ner p , s} $ and $a_{\ner p , s}$, the effective hamiltonian (\ref{3.18}) assumes the diagonal form 
 \be
H  \simeq   \sum_{\ner {p} , s } \varepsilon ({\ner {p}})   \,  a^\dagger_{\ner {p} , s }  a_{\ner {p} s }  
+ {\cal E}_0 \; ,  
\label{3.24}
\ee 
where 
\be
{\cal E}_0 =  \sum_{\ner {p} } \left [ w(p) - \varepsilon ({\ner {p}}) -  \Delta_{\ner {p}  }  X_{\ner {p}}   \right ]  + E_F(V , \mu )  \; .  
\label{3.25}
\ee
The operators $a^\dagger_{\ner p , s} $ and $a_{\ner p , s}$  verify  
\bea
\{ a_{\ner {p} , s} \, , a^\dagger_{\ner {q} , r } \} &=& \delta_{sr} {h^3\over V} \delta (\ner {p} - \ner {q})\quad ,\nonumber \\  
\{ a_{\ner {p} , s} \, , a_{\ner {q}, r } \} &=& 0 = \{ a^\dagger_{\ner {p} , s} \, , a^\dagger_{\ner {q} , r } \} \; ,  
\label{3.26}
\eea
and represent the creation and annihilation operators for  quasi-particles (the so-called bogolons) with  dispersion relation $\varepsilon (\ner p)$.   The grand partition function (\ref{3.3}) is given by 
\be
{\cal Q}(T,V,\mu ) =   {\rm Tr} \left [ e^{- H  / kT  } \right ]  = e^{-{\cal E}_0 / kT }\, \prod_{{\ner {p} } , s}  \left [ 1 + e^{- \varepsilon (\ner p)   / kT } \right ] \; ,  
\label{3.27}
\ee
and then the grand potential $\Omega$ for the system of conducting electrons turns out to be 
\be
\Omega (T , V , \mu ) = - kT  \sum_{\ner {p} , s } \ln  \left [ 1 + e^{- \varepsilon ( {\ner {p} } )    / kT } \right ] +  {\cal E}_0 \; .   
\label{3.28}
\ee
The introduction of  Step~1 clarifies the nature of the bogolons and the resulting formalism is consistent with the vanishing of the bogolon chemical potential. Note that,  if the Bogoliubov transformation  is directly applied to  the creation and annihilation operators $ b^\dagger_{\ner {p} , s }$ and $  b_{\ner {p} , s }$ of the electrons, one finds sign ambiguities \cite{RIC} in the energy spectrum.  The method  that has been presented here has the advantage of avoiding  these ambiguities. By the way,  the correct solution of these ambiguities is rather nontrivial. Because,  if the electron effective energy $\eta (p)$ takes the place of $w(p)$,  the thermodynamic consistency requires that the right sign of the  dispersion relation should be $+ \varepsilon (\ner p)$ when $|\ner p | > p_0$ and  $- \varepsilon (\ner p)$ when $|\ner p | < p_0$.   The correct value (\ref{3.25}) of the   energy ${\cal E}_0$   of the vacuum of the quasi-particles  is a  fundamental ingredient in the following discussions in  Section~{\ref{sec:4}} and Section~{\ref{sec:5}}. Group properties of the quasi-particles vacuum and features of the solution of the gap equation have been discussed, for instance, in Ref.\cite{T,BF,HHS}. 
 
In order to complete the derivation of $\Omega $, one has to impose the statistical self-consistency of the approximation (\ref{3.8}), in which the mean values (\ref{3.6}) have been used. Let us briefly recall the method which is usually presented in literature  for this purpose.  The mean values (\ref{3.6}) can be related  with the mean values of the corresponding operators defined in terms of $a^\dagger_{\ner p , s}$ and $a_{\ner p , s}$. Since the quasi-particles created and annihilated by $a^\dagger_{\ner p , s}$ and $a_{\ner p , s}$ behave as free particles, the only nontrivial mean value is given by $\langle a^\dagger_{\ner p , s} a_{\ner p , s} \rangle = n_{\ner p}$, where the quasi-particle distribution $n_{\ner p}$ is shown in equation (\ref{2.5}). In this way, by using the definition (\ref{3.9}), one can obtain  the gap equation \cite{BCS,L3,RIC,PN,GO}
\be
1  = {U_0 \over V} {\sum_{\ner {p} }}^\prime \, { 1\over 2 \varepsilon ({\ner {p}}) } \; {\rm {Th}} \left [ \varepsilon ({\ner {p}}  ) /2 kT \right ]\; ,    
\label{3.29}
\ee
where ${\sum_{\ner {p} }}^\prime$ denotes the integral in the neighbourhood  of the surface of the Fermi sphere which is specified by the condition  $ \big |(\ner p^2 / 2 m) - \mu \big | \leq \hbar \omega_D$. In the next section, an alternative derivation of the gap equation will be produced.  

\section{Thermodynamic consistency and the gap equation}
\label{sec:4}

From the definition (\ref{3.9}) and the experimental fact that the superconducting electrons belong to  a neighbourhood of the Fermi surface, where $U_{\ner {p} , \ner {q}}\simeq - U_0/V$,  it follows that one can put 
\be
\Delta_p  = \left  \{ \begin{array}  {l@{ ~ } l}  
  \Delta   & \hbox{ ~for ~ }    \big |(\ner p^2 / 2 m) - \mu \big | \leq \hbar \omega_D\, ;   \\  ~ & ~ \\  0 &  \hbox{ ~otherwise;  }  
\end{array} \right. 
\label{4.1}
\ee
and then 
\be
- \sum_{\ner p} \Delta_p  X_p  = {V\over U_0} \Delta^2 \; .   
\label{4.2}
\ee
Let us now consider the expression (\ref{3.28}) and let us assume, for the moment, that the value $\Delta $ of the gap is a free undeterminate which is not specified by the gap equation; one has   
\be
\Omega (T,V, \mu ) = 
- 2 kT  \sum_{\ner {p} } \ln  \left [ 1 + e^{- \varepsilon ( {\ner {p} } )    / kT } \, \right ] + {\cal E}_0 \; , 
\label{4.3}
\ee
where
\be
{\cal E}_0 =  \sum_{\ner p} \,  \left [ \, \left |  (\ner p^2 / 2 m) - \mu \right | - \varepsilon (\ner p)  \right ] + {V\over U_0} \Delta^2 + E_F(V , \mu)  \; .     
\label{4.4}
\ee
Let us denote by ${\cal B}$ the neighbourhood of the Fermi surface  which is defined by the condition $ \big |(\ner p^2 / 2 m) - \mu \big | \leq \hbar \omega_D$. Then   
 \be
\varepsilon (\ner p ) = \left  \{ \begin{array}  {l@{ ~ } l}  
 \left [   \left |  (\ner p^2 / 2 m) - \mu \right |^2 +  \Delta^2 \right ]^{1/2} & \hbox{ ~for ~ }    \ner p \in {\cal B} ;   \\  ~ & ~ \\   \left |  (\ner p^2 / 2 m) - \mu \right | &  \hbox{ ~otherwise.  }  
\end{array} \right. 
\label{4.5}
\ee
Since $\Omega $ is the thermodynamic potential associated with the variables $(T,V,\mu )$, for fixed $(T,V,\mu )$  the system tends to minimize $\Omega$. So, in the equilibrium state, the value $\widetilde \Delta $ of the gap can be obtained by imposing the variational condition 
\be
\left ( {\der \Omega \over \der \Delta } \right )_{T,V,\mu }\, \bigg |_{\Delta = \widetilde \Delta} = 0 \; . 
\label{4.6}
\ee
Since 
 \be
{\der \varepsilon ( \ner p )  \over \der \Delta } = \left  \{ \begin{array}  {l@{ ~ } l}  
 \Delta /  \varepsilon (\ner p )  & \hbox{ ~when ~ }    \big |(\ner p^2 / 2 m) - \mu \big | \leq \hbar \omega_D\, ;   \\  ~ & ~ \\   0 &  \hbox{ ~otherwise,   }  
\end{array} \right. 
\label{4.7}
\ee
one finds 
 \be
{\der \over \der \Delta } \left \{ - 2 kT  \sum_{\ner {p}  } \ln  \left [ 1 + e^{- E_{\ner {p} }   / kT } \right ] \right \} =  \Delta  {\sum_{\ner p}}^\prime {1\over \varepsilon (\ner p)} \,  {2\over e^{\varepsilon (\ner p) /kT} + 1} \; . 
\label{4.8}
\ee
Moreover 
\be 
{\der \over \der \Delta }{\sum_{\ner p}} \left [ \left | ( {\ner p}^2 / 2 m) - \mu \right | - \varepsilon (\ner p )  \right ] = - \Delta {\sum_{\ner p}}^\prime {1 \over \varepsilon (\ner p) } \; , 
\label{4.9}
\ee
and 
\be
{\der \over \der \Delta } \left (  {V\over U_0} \Delta^2 \right ) =  \Delta {2V\over U_0} \; . 
\label{4.10}
\ee
Equations (\ref{4.8})-(\ref{4.10}) imply 
\be
{\der \Omega  \over \der \Delta }  = \Delta  \left \{ {\sum_{\ner p}}^\prime {1\over \varepsilon (\ner p)} \left [ {2\over e^{\varepsilon (\ner p) /kT} + 1} -1 \right ] + {2V\over U_0} \right \} \; . 
\label{4.11}
\ee
By using the identity 
\be
{2\over e^{\varepsilon ({\ner {p}}) / kT  } + 1     } - 1 =  - {\rm {Th}} (\varepsilon ({\ner {p}}) /2 kT )  \; , 
\label{4.12}
\ee
 the minimum condition (\ref{4.6})  turns out to be equivalent to  (apart from the trivial solution) 
\be
1  = {U_0 \over V} {\sum_{\ner {p} }}^\prime \, { 1\over 2 \varepsilon ({\ner {p}}) } \; {\rm {Th}} (\varepsilon ({\ner {p}}) /2 kT )\; ,    
\label{4.13}
\ee
which coincides precisely with the gap equation (\ref{3.29}). In what follows, the value of the gap
satisfying equation (\ref{4.13}) will be simply denoted by $\Delta = \Delta (T, \mu ) $.  

To sum up, the grand potential $\Omega$ for the electrons system takes the form 
\be
\Omega (T, V, \mu) = F_q (T,V  ; \Delta (T, \mu)) + {\cal E}_0 (V, \mu ; \Delta (T, \mu)) \; ,  
\label{4.14}
\ee
where $F_q$ denotes the free energy for the gas of quasi-particles, with dispersion relation $\varepsilon (\ner p)$ shown in equation (\ref{4.5}),  and ${\cal E}_0$ of equation  (\ref{4.4})  represents the vacuum energy of the quasi-particles. The value $\Delta (T, \mu)$ of the gap corresponds to a minimum of $\, \Omega \, $ for fixed $(T,V,\mu)$.

Let us now consider the issue of the thermodynamic consistency of the result (\ref{4.14}).   The starting expression (\ref{3.3}) is thermodynamically  consistent, because the   dependence of the grand partition function on the macroscopic variables $(T,V,\mu )$ is the standard dependence of statistical physics. However, when the dependence of the gap on the variables $T$ and $\mu $ has been introduced, $\Delta = \Delta (T, \mu)$, the standard rules of statistical physics have been modified.  Now, the dependence of $\Omega $ on the variables $(T, V , \mu )$ consists of two parts:

\begin{itemize}
\item the (thermodynamically consistent)  explicit  dependence on $(T, V , \mu )$ displayed in expressions (\ref{4.3})-(\ref{4.5}); 
\item the indirect dependence on $(T , \mu )$ through $\Delta = \Delta (T, \mu ) $. 
\end{itemize}

\noindent Let us distinguish these two possibilities by means of the notation 
\be
\Omega = \Omega (T, V , \mu ; \Delta (T, \mu )) \; . 
\label{4.15}
\ee
When computing the derivatives of $\Omega$ one finds, for instance,   
\be
{\der \Omega \over \der T} = \left ( {\der \Omega \over \der T} \right )_{\! \Delta} + \left ( {\der \Omega \over \der \Delta} \right )_{\! T} \, {\der \Delta \over \der T}  \; . 
\label{4.16}
\ee
But since the value of the gap minimizes $\Omega$, the  term $( \der \Omega / \der \Delta )_T$ vanishes, and then the derivative  $\der \Delta / \der T $ does not contribute to $\der \Omega / \der T$.  Therefore,  only the thermodynamically consistent dependence of $\Omega $ on $(T, V, \mu )$ really contributes to the derivates. This means that the Maxwell relations associated with $\Omega$ are satisfied.  

Let us now consider a generic thermodynamic potential which is derived from $\Omega $ by means of appropriate Legendre transformations.  As far as  the variable $\Delta$ is concerned, the so-called {\it theorem of small increments} \cite{L4} takes the form  
\be
0 = \left ( {\der \Omega \over \der \Delta} \right )_{T,V, \mu } = \left ( {\der F \over \der \Delta} \right )_{T,V,N} = \left ( {\der U \over \der \Delta} \right )_{V,S,N} = \cdots \; , 
\label{4.17}
\ee
and states that $\Delta $ really corresponds to a stationary point of each potential (provided  the corresponding set of thermodynamic variables is kept fixed). Consequently,  when constructing  the thermodynamic relations by means of the first derivatives of the potentials, $\Delta$ effectively behaves like a constant term, and does not alter the Maxwell relations.  Thus,  because of the validity of the gap equation,  the quasi-particles description of the systems,  which is given ---in the low temperature limit--- by the grand potential (\ref{4.14}),   is perfectly consistent with the  validity of the standard thermodynamic relations. 

This  example of superconducting electrons  
is in agreement with the Landau principle, which states that the value of any order parameter entering a given thermodynamic potential actually corresponds to a stationary point of the same potential, with the appropriate thermodynamic variables held  fixed. 

\section{Thermodynamic potentials}
\label{sec:5}

Quite often, the complete expression (\ref{4.14}) of the grand potential and, in particular, the complete expression (\ref{4.4})   of the energy ${\cal E}_0$ are not explicitly displayed  in literature.    In certain  cases, the combination $ \sum_{\ner {p} } \left [ w(p) - \varepsilon ({\ner {p}}) -  \Delta_{\ner {p}  }  X_{\ner {p}}   \right ]$ is reported, however  the  term $E_F(V, \mu )$ is missing. Consequently, in these cases the partial derivatives of $\Omega $ with respect to $V$ and $\mu $ are invalidated and thus the deduction of the thermodynamic potentials as functions of the variables $(T,V,N)$ becomes problematic. Usually, the explicit derivation of  the general dependence of the superconducting thermodynamic potentials  on $(T,V,N)$ represents a rather laborious task. So, let us now derive from equation (\ref{4.14}) a few thermodynamic potentials as functions of $(T,V,N)$  when $T<T_c $ in the low temperature limit. Let us compute, in particular,   how the chemical potential  gets modified by the superconducting phase transition  and how $\mu $ depends on the electron density. The dependence of $\mu$  on $(T,V,N)$  is indeed  the fundamental ingredient for the   determination of the consistent thermodynamic potentials.    

In the $T\rightarrow 0$ limit, the quasi-particles distribution is mainly concentrated around the minimum of the energy spectrum. The energy (\ref{3.21}) of a single quasi-particle  can be written as  $\varepsilon  =  \varepsilon_w =  \sqrt {w^2 + \Delta^2\, }$ and,  in a neighbourhood of the minimum $w=0$, it    can be approximated as 
\be
  \varepsilon_w =  \sqrt {w^2 + \Delta^2\, } \simeq \Delta  + { w^2 \over 2 \Delta }   \; .    
\label{5.1}
\ee
Accordingly, in the computations of $F_q $ and  of ${\cal E}_0$  the integral in momentum space is effectively dominated by the integration in a  neighbourhood of the surface of the Fermi sphere and it can be written as  
\be 
\sum_{\ner p} \, \rightarrow \, {V (2m)^{3/2} \over 2 \pi^2 \hbar^3} \sqrt \mu \int dw \; . 
\label{5.2}
\ee
The grand potential (\ref{4.14}) reads 
\bea
\Omega (T , V , \mu ) &\simeq& -  V kT  {(2 m)^{3/2} \over  \pi^2 \hbar^3} \mu^{1/2}\! 
  \int_0^\infty \! dw \,  \ln \left [ 1 + e^{- \varepsilon_w /kT} \right ] \nonumber \\ 
&& + V  {(2 m)^{3/2} \over 2 \pi^2 \hbar^3} \mu^{1/2}  \int_{0}^{ \hbar \omega_D} dw \left [ w - \varepsilon_w \right ]   \nonumber \\ 
&& + {V \over U_0} \Delta^2 
 - {2 V  (2m)^{3/2}\over 15 \pi^2 \hbar^3} \mu^{5/2} 
 \; .  
\label{5.3}
\eea 
In the thermodynamic limit, the statistical mean value $N$ of the number of electrons  is specified by  the relation $N = - \der \Omega / \der \mu $, which takes the form  
\bea 
{3 \pi^2 \hbar^3 \over (2 m)^{3/2}} {N \over V}  &\simeq&  \mu^{3/2} + 
{3 kT \over 2 \mu^{1/2} }\int_0^\infty dw \,  \ln \left [ 1 + e^{- \varepsilon_w /kT} \right ]   \nonumber \\ 
&& - {3 \over 4 \mu^{1/2}}   \int_{0}^{ \hbar \omega_D} dw \left [ w - \varepsilon_w \right ]   \; .   
\label{5.4}
\eea
In the low temperature limit one has $\Delta / \varepsilon_F \ll 1$ and $(kT / \varepsilon_F) \ll 1$, where $\varepsilon_F$  denotes the Fermi energy 
\be
\varepsilon_F =  { \hbar^2 \over 2 m } \left ( {N\over V} \right )^{2/3} \left ( {3 \pi^2 } \right )^{2/3}  \; .  
\label{5.5}
\ee 
So, the chemical potential can be determined perturbatively   in powers of $(kT / \varepsilon_F) \ll 1$ and $\Delta / \varepsilon_F\ll 1$ from equation (\ref{5.4}); to lowest non-trivial orders one finds   
\be
\mu \simeq \varepsilon_F    -  { kT \over \varepsilon_F }\int_0^\infty dw \,  \ln \left [ 1 + e^{- \varepsilon_w /kT} \right ] +  {1 \over 2 \varepsilon_F}   \int_{0}^{ \hbar \omega_D} dw \left [ w - \varepsilon_w \right ] \; .  
\label{5.6}
\ee
This relation gives the desired expression for  the chemical potential. It should be noted that, in addition to the quasi-particles contributions to $\mu$,  the last term on the right-hand-side of  equation (\ref{5.6}) describes precisely the effects of the quasi-particles vacuum rearrangement  ---that takes place at the critical point---  on the chemical potential.    The free energy $F$ is given by  $F = \Omega + N \mu \, $, 
\bea
F(T, V , N ) &\simeq& -  { 3 N kT \over \varepsilon_F } \int_0^\infty dw \,  \ln \left [ 1 + e^{- \varepsilon_w /kT} \right ] + {3\over 5} N \varepsilon_F \nonumber \\
&& \; \; + {3N \over 2 \varepsilon_F}   \int_{0}^{ \hbar \omega_D} dw \left [ w - \varepsilon_w \right ]  + {V\over U_0} \Delta^2 \; . 
\label{5.7}
\eea
The entropy $S = - \der F / \der T$ turns out to be 
\bea
S(T, V , N ) &\simeq&  { 3 N k \over \varepsilon_F } \int_0^\infty dw \,  \ln \left [ 1 + e^{- \varepsilon_w /kT} \right ] \nonumber \\ 
&&  + { 3 N  \over T \varepsilon_F } \int_0^\infty dw \,  {\varepsilon_w \over  e^{ \varepsilon_w /kT} + 1 }  \; ,  
\label{5.8}
\eea
and coincides with the entropy of the gas of quasi-particles, as it must be. Finally, the internal energy  $U = F + T S$ is given by    
\bea
U(T, V , N ) &\simeq&  { 3 N  \over \varepsilon_F } \int_0^\infty dw \,  {\varepsilon_w \over  e^{ \varepsilon_w /kT} + 1 }  + {3 \over 5} N\varepsilon_F  \nonumber \\
&&  {\hskip -0.5 cm} + { 3 N  \over 2\varepsilon_F } \int_{0}^{ \hbar \omega_D} dw \left [ w - \varepsilon_w \right ]  + {V\over U_0} \Delta^2\; .  
\label{5.9}
\eea
The consequences of expressions (\ref{5.3})-(\ref{5.9}) are in  agreement with the known   results \cite{BCS,L3,TK,RIC,AGD,AZA} on superconductivity.  As a check, let us compute the condensation energy. The energy of the quasi-particles gas can be obtained by means of the gaussian approximation 
\be
{ 3 N  \over \varepsilon_F } \int_0^\infty dw \,  {\varepsilon_w \over  e^{ \varepsilon_w /kT} + 1 } \simeq {3N \Delta \over 2 \varepsilon_F} \, e^{- \Delta / kT} \, \sqrt {2 \pi \Delta kT} \, ; 
\label{5.10}
\ee
this energy contribution vanishes in the $T \rightarrow 0 $ limit.  Since $\Delta / \hbar \omega_D \ll 1$, one finds 
\be
\int_{0}^{ \hbar \omega_D} dw \left [ w - \varepsilon_w \right ]  \simeq - \quarto \Delta^2 + \mezzo \Delta^2 \, \ln (\Delta / 2 \hbar \omega_D)\; . 
\label{5.11}
\ee 
In the $T \rightarrow 0 $ limit, the value of $\Delta$ is specified by  the gap equation (\ref{4.13}); in particular, when $(3NU_0 / 4 V \varepsilon_F) \ll 1$, one obtains   \cite{BCS,L3,TK}
\be
(\Delta / 2 \hbar \omega_D) \simeq e^{- (4V \varepsilon_F / 3 N U_0)} \; .   
\label{5.12}
\ee
By means of relation (\ref{5.12}), one gets 
\be
 { 3 N  \over 2\varepsilon_F } \int_{0}^{ \hbar \omega_D} dw \left [ w - \varepsilon_w \right ]  + {V\over U_0} \Delta^2 \simeq - {3 N \Delta^2 \over 8 \varepsilon_F} \; . 
 \label{5.13}
 \ee
Within the considered approximations, the value $E_C$ of the condensation energy  is obtained by subtracting the energy of a free fermions gas at $T=0$ from the system energy $U( T =0 , V, N) $. Since  the energy of a free electrons gas in the Fermi sphere is equal to $(3/5) N \varepsilon_F$, the condensation energy $E_C$ turns out to be 
 \be 
 E_C= U (T=0, V , N) - {3\over 5} N \varepsilon_F = -  {3 N \Delta^2 (T=0) \over 8 \varepsilon_F} \; ,    
\label{5.14}
 \ee
 which is in agreement with the results of the alternative derivations presented in \cite{BCS,L3,TK}. 

\section{Summary and conclusions} 
\label{sec:6}

In the Bogoliubov-Valatin approach, it has been shown that the relevant variables ---that must be considered for the computation of the thermodynamic potentials--- are associated with the energy fluctuations of the system around the Fermi sphere configuration. By means of a Bogoliubov transformation in  these variables one gets the expression of the 
quasi-particles vacuum energy, which is the fundamental ingredient in the construction of the  grand potential $\Omega (T, V , \mu)$ for the superconducting electrons.  The value of the gap corresponds to a minimum of $\Omega$ and the gap equation precisely represents the stationarity condition.   When the gap equation is satisfied, $\Omega$ and the thermodynamic potentials that are derived from $\Omega $ satisfy the Maxwell relations. 
The  expressions of the thermodynamic potentials as functions of  the $(T,V,N)$ variables have been derived.  The dependence of the chemical potential on the electron density and the effects of the 
superconducting phase transition on the chemical potential have been computed in the low energy limit. As a final check, a derivation of the condensation energy from $\Omega (T,V, \mu )$ has been presented. 

\vskip 0.7 truecm 

\noindent {\large \bf Acknowledgments.} I wish to thank E. Vicari for discussions.

\end{document}